\documentclass[preprint,aps]{revtex4}
\include{epsf}   
\include{rotate}
\include{amsmath} 
\begin{document} 
\title{\bf Sandpile models and random walkers on finite lattices} 
\author{Yehiel Shilo and Ofer Biham} 
\affiliation{ 
Racah Institute of Physics,  
The Hebrew University,  
Jerusalem 91904,  
Israel} 
 
\begin{abstract} 
Abelian sandpile models, both deterministic, 
such as the Bak, Tang, Wiesenfeld (BTW) model 
[P. Bak, C. Tang and K. Wiesenfeld, 
Phys. Rev. Lett. {\bf 59}, 381 (1987)], 
and stochastic, such as the Manna model 
[S.S. Manna, J. Phys. A {\bf 24}, L363 (1991)], 
are studied on finite square lattices with open boundaries. 
The avalanche size distribution $P_L(n)$
is calculated for a range of  
system sizes, $L$. The first few moments  of this distribution  
are evaluated numerically and their dependence on the system  
size is examined. The sandpile models are conservative in  
the sense that grains are conserved in the bulk and  
can leave the system only through the boundaries. 
It is shown that the conservation law provides an interesting   
connection between sandpile models 
and random walk models. 
Using this connection, it is shown that  
the average avalanche sizes, 
$\langle n \rangle_L$,
for the BTW and  
the Manna models are equal to each other, 
and both are equal to 
the average path-length of a random walker starting from a 
random initial site on the same lattice of size $L$. 
This is in spite of  
the fact that  
sandpile models with 
deterministic (BTW) 
and stochastic (Manna) 
toppling rules
exhibit different critical exponents, indicating that they
belong to different universality classes. 
\end{abstract} 
 
\pacs{PACS: 05.70.Jk,05.40.Fb,05.70.Ln} 

\maketitle 
 
\section{Introduction} 
 
Sandpile models have been studied extensively in the past 
fifteen years as a paradigm of  
self organized criticality (SOC) 
\cite{Bak1987,Bak1988,Tang1988}.  
SOC provides a useful framework for the analysis of driven 
nonequilibrium systems which dynamically evolve into a critical state. 
At the critical state these systems exhibit  
avalanche dynamics with 
long-range spatial and  
temporal correlations, which resembles the behavior at equilibrium  
critical points.  
In sandpile models,  
defined on a lattice, grains are deposited  
randomly until the height at some site exceeds the threshold, 
thus becoming unstable. 
The unstable site is toppled and grains are distributed between its  
nearest neighbors, which may become unstable too, resulting in 
an avalanche. 
These models were found to be self-driven into a critical state 
in which the avalanche sizes follow 
a power-law distribution. 
The critical state, which can be characterized by various  
critical exponents and scaling functions,  
was studied  
using both theoretical 
\cite{Dhar1989,Hwa1989,Dhar1990,Carlson1990,Grinstein1991,Paczuski1996,Vespignani1998,Vanderzande2001} 
and numerical approaches 
\cite{Kadanoff1989,Grassberger1990,Manna1990,Christensen1991,Christensen1993,Paczuski1997,Lubeck1997b,Lubeck1997a,Ktitarev2000}. 
These studies stimulated an effort to examine the utility
of the SOC framework to the understanding of empirical phenomena
such as earthquakes
avalanches in granular flow
and mass extinctions
\cite{Bak1996}.

To examine the dependence of the critical state on  
various properties of the models, 
different sandpile models  
have been introduced. 
These include the stochastic model introduced by  Manna 
\cite{Manna1991a}. 
The issue of universality has been studied.
Analytical studies 
\cite{Pietronero1994,Vespignani1995}
and numerical simulations  
\cite{Chessa1999} 
indicated  
that the Manna model, which is stochastic, 
belongs to the universality class 
of the original model introduced by  
Bak, Tang and Wiesenfeld (BTW) which is 
deterministic
(namely, has a deterministic toppling rule).
However, 
numerical simulations using an extended set of critical exponents 
showed that deterministic and stochastic models 
exhibit different scaling properties and thus belong to different 
universality classes 
\cite{Ben-Hur1996,Milshtein1998,Biham2001}.  
Further support for this result was obtained  
using multifractal analysis 
\cite{Tebaldi1999}, 
moment analysis 
\cite{Lubeck2000} 
as well as studies of sandpile models as closed systems 
\cite{Dickman1998,Dickman1999,Vespignani2000}. 
The crossover between the two classes was also studied
\cite{Lubeck2000b}.
In the case of directed models it was shown analytically
that deterministic and stochastic models belong to 
different universality classes
\cite{Dhar1989,Paczuski2000,Kloster2001}.
 
In this paper we present a connection between abelian sandpile 
models and random walkers on finite lattices,  
which is  a result of the conservation laws. 
In the sandpile models each avalanche starts with  
the addition of one grain.  
The models are conservative in the sense that grains are  
conserved in the bulk of the system 
and can leave it only through the boundaries. 
It is thus clear that under steady state conditions,  
the average number of grains leaving the system 
per avalanche is also one. 
Here, the avalanche size is defined
as the number of hops of grains that are 
toppled from unstable sites during an avalanche.
The avalanche size is thus equal to the number of
toppling events (or relaxations) of unstable sites 
during the avalanche times the number of grains that
topple in each event of this type. 
We show that in both the BTW and the Manna models,  
each grain moves like a random walker, 
starting at the site in which it was deposited, until it 
falls off the boundaries. 
Using these features we show that
the average avalanche size 
is the same for the two models. 
Moreover, it is  equal to the average path length of a 
random walker starting from a random site 
on the same lattice, until it falls off the edge.

In order to demonstrate these properties we examine the avalanche  
size distribution of the Abelian sandpile models and the 
distribution of path length of a random walker on finite square lattices. 
The average path length of a random 
walker on a lattice of size $L$
is calculated exactly using  
a method proposed by Walsh and 
Kozak \cite{Walsh1981,Walsh1982}. 
The entire distribution  of the path 
lengths of  random walkers starting at random 
sites on the finite lattice is also 
calculated using a related method proposed by Soler \cite{Soler1982}. 
The results are also compared to direct numerical simulation of the random  
walk. The avalanche size distribution of the Abelian sandpile 
models is obtained  from direct numerical simulations as  
well as from an exact formula 
introduced by Dhar \cite{Dhar1990}. 
 
The paper is organized as follows. The sandpile models are described in  
Sec. II. The distributions of path lengths of 
random walkers on finite lattices  and their averages are studied in Sec. III. 
The connection between sandpile models and random walkers is 
examined in Sec. IV. The simulations and results are given in Sec. V, 
followed by  a discussion in Sec. VI and a summery in Sec. VII. 
 
\section{Sandpile Models} 
 
 
Consider a sandpile model on a $d$-dimensional cubic  
lattice of linear size $L$. 
Each site {\bf i} is assigned a dynamic variable $E({\bf i})$ which  
represents some physical quantity such as energy, grain density,  
stress, etc.  
A  configuration 
$\left \{E({\bf i}) \right \}$ is called {\it stable} if 
for all sites $E({\bf i}) < E_c$,  where $E_c$ is a threshold value. 
The evolution between stable configurations is by the following  
rules: 
(i) Adding energy. 
Given a stable configuration $\{E({\bf j})\}$ we select a  
site ${\bf i}$ at random and increase $E({\bf i})$ by some amount  
$\delta E$. 
When an unstable configuration is reached rule (ii) is applied. 
(ii) Relaxation (or toppling) rule. 
If $E({\bf i}) \geq$  
$E_c$,    
relaxation takes place and energy is distributed in the following way: 
 
\begin{eqnarray}\label{eq:vector} 
\label{def} 
E({\bf i}) & \rightarrow & E({\bf i}) - \sum_{\bf e}\Delta E({\bf e})  \nonumber \\				 
E({\bf i}+{\bf e}) & \rightarrow & E({\bf i}+{\bf e})+ 
\Delta E({\bf e}),  
\end{eqnarray} 
 
\noindent 
where ${\bf e}$ are a set of vectors from the site ${\bf i}$ 
to some neighbors.  
As a result of the relaxation, $E({\bf i}+{\bf e})$ for one or more of 
the neighbors may reach or exceed the threshold $E_c$.   
The relaxation rule is then applied until a stable  
configuration is obtained. 
The resulting sequence of topplings is an avalanche which propagates  
through the lattice. 
 
Avalanches can be characterized by their 
size. 
The size $s$ of an avalanche is the total number of toppling
events that occurred during the course of the avalanche.
In the models studied here the number of grains
that topple from an unstable site is
$E_c = \sum_{\bf e}\Delta E({\bf e})$.
Throughout the rest of the paper we will
denote the avalanche size by

\begin{equation}
n = E_c \cdot s,
\end{equation}

\noindent
namely, by the number of hops of individual grains that take
place during the avalanche.
This will allow us to consider models with different values
of $E_c$ on a common footing.

The avalanche size distribution is denoted by
$P_L(n)$, $n=0,1,\dots$, namely, the probability
of a randomly chosen avalanche to be of size $n$.
The normalization condition is given by

\begin{equation}
\sum_{n=0}^{\infty} P_L(n) = 1.
\label{eq:normalization} 
\end{equation} 

\noindent
Numerical simulations show that the avalanche  
size distribution for a lattice of size $L$ 
has the power-law form:
 
\begin{eqnarray}
\label{eq:power} 
{ 
P_L(n) \sim n^{-\tau_L}, \ \ \ n = 1,2,\dots. 
} 
\end{eqnarray} 

\noindent
For some of the models
the results for $\tau_L$ exhibit a significant dependence
on the system size.
The critical exponents $\tau_L$ 
turn out to depend on the 
vector $\Delta E$ to be termed {\it relaxation vector}.
For a square lattice with relaxation to nearest neighbors it is of  
the form $\Delta E=(E_N, E_E, E_S, E_W)$,   
where $E_N, E_E,  E_S$ and $E_W$ are 
the amounts transferred to the northern,  eastern,   
southern and western nearest neighbors respectively.  
The average avalanche size on a lattice of size $L$
is given by

\begin{equation}
\langle n \rangle_L = \sum_{n=0}^{\infty} n P_L(n).
\end{equation}

\noindent
The sandpile models considered in this paper are conservative in the sense 
that the grains  are conserved  in the bulk and can leave  
the system only through  the open boundaries. 
When an avalanche reaches a boundary 
site, some energy is transferred out of the system (namely, dissipation 
takes place at the boundaries). 
The critical state is reached spontaneously  
in the limit in which the 
random addition of energy (or drive)  
is infinitely slow (practically it means that 
the next energy 
unit is added only after the previous avalanche is completed). 
This state is 
characterized by a power-law distribution of avalanche 
 sizes [Eq.~(\ref{eq:power})]. 
In the critical state the added energy  
$\delta E$ per avalanche,  
is balanced, on average, by the energy that flows out through the boundaries. 
Therefore, the average amount of energy leaving the system per avalanche is  
$\delta E$. 
 
In the BTW model,  
$E_c=4$, $\delta E=1$ and $\Delta E=(1, 1, 1, 1)$.  
Since $\Delta E$ is a constant, this model is clearly deterministic. 
Note that since $\Delta E$ is independent of 
$E({\bf i})$,  
if an active  
site with   
$E({\bf i}) > E_c$  
is toppled, it remains non-empty after  
the toppling event had occurred.  
A useful way to analyze the BTW model is by its toppling matrix $\Delta$, 
which for an  $L \times L$  
 lattice is a matrix of size  
$L^2 \times L^2$. 
Consider a pair of sites  
${\bf i }=(i_{x},i_{y})$, 
${\bf j }=(j_{x},j_{y})$ 
and denote ${ i } ={ Li_{x}+i_{y}}$ 
 ,  ${ j } = { Lj_{x}+j_{y}}$ 
 where $ 0 \leq i_{x},i_{y},j_{x},j_{y} \leq L-1$. 
 The matrix element $\Delta_{i,j}$ where $ {i} \neq {j}$ 
  is the number 
 of grains given to site $\bf{j}$ when site $\bf{i}$ topples (up to  
a minus sign). The number 
 of grains leaving  site $\bf{i}$ in such event is  
given by the diagonal element $\Delta_{i,i}$. Therefore, the  
toppling  matrix is : 
 
\begin{eqnarray}\label{eq:delta} 
\Delta_{i,j} = \left\{ \begin{array}{ll}		 
		 4  &  \ \ {\bf i} = {\bf j} \\ 
		-1 & \ \ {\bf i} \ {\rm and} \ {\bf j} \ \rm{are  
				\ nearest \  neighbor \ sites}\\ 
		0 & \ \ \rm{otherwise.} \\ 
		\end{array} 
\right. 
\end{eqnarray} 
 
\noindent 
Consider the toppling  of a given site {\bf i}. 
It can 
be described by: 

\begin{eqnarray} 
{ 
E({\bf j } )  \rightarrow  E({\bf j } ) - \Delta_{ij},  
} 
\end{eqnarray} 
 
\noindent 
for all sites ${\bf j}$. 
 
In the class of stochastic sandpile models, 
introduced by Manna, 
a set of neighbors is randomly chosen  
for relaxation 
\cite{Manna1991a} 
once a site becomes unstable. 
Such models can be specified by a set of relaxation vectors, each  
vector is assigned with a probability for its appearance. 
There are several models in this class.
One of them is a two-state model with $E_{c}=2$ 
and two  
relaxation vectors (1,0,1,0) and (0,1,0,1) each one applied with a  
probability of $1/2$ \cite{Dhar1999b}. 
Another two-state model  
includes six relaxation   
vectors, namely 
(1,1,0,0), (1,0,0,1),  
(0,1,1,0), (0,0,1,1), (1,0,1,0) and (0,1,0,1),    
each one applied with a  
probability of $1/6$ \cite{Ben-Hur1996}. 
In this paper we consider a two-state  model in which each of the two 
grains of 
an unstable site is toppled randomly to one of the four neighbors 
(with probability 1/4 to each direction). There is no correlation  
between the directions picked for these two grains. The set of 
relaxation vectors includes all the 10 possible vectors $\Delta E$ 
of integer components for which $E_{N} + E_{E}+E_{W}+E_{S} = 2$.  
Each of the six vectors of the previous model appears with probability 2/16,  
while each of the four vectors: 
(2,0,0,0), (0,2,0,0), (0,0,2,0) and (0,0,0,2),  
appears with probability 1/16. This model will be called the 
unrestricted two-state Manna model. 

The average avalanche size
$\langle n \rangle$ 
for the BTW model on a lattice of size
$L$ was calculated exactly by Dhar
\cite{Dhar1990}.
He showed that
the matrix element 
$\Delta ^{-1}_{i,j}$ represents the average number of toppling 
events taking place at site {\bf j} 
when a grain starts an avalanche after being deposited at  
site {\bf i}. 
By summing all the elements of $\Delta ^{-1}$, using  
the eigenvectors and eigenvalues  
of $\Delta$ \cite{Dhar1995}  
it was found that
 
\begin{eqnarray} 
\label{eq:Dhar1} 
\langle n \rangle   =  \frac{1}{L^{2}(L+1)^2}
\sum_{k,l}
{ {\cot^{2} \theta_k 
\cot^{2} \theta_l}
\over  
{\sin^{2} \theta_k +
\sin^{2} \theta_l} },
\end{eqnarray} 
 
\noindent
where 

\begin{equation}
\theta_m = \frac{\pi m}{2(L+1)}, 
\end{equation}

\noindent
for any integer $m$,
and the summation over $k,l$ 
is over all odd integers $1\leq k,l \leq L$. 
The dependence of 
$\langle n \rangle$  
on the system size was found to be
$\langle n \rangle \sim L^{2}$.  
This analysis was recently extended to dissipative abelian models
\cite{Tsuchiya2000}.

\section{Random walks on finite lattices} 
  
Consider a random walker on a 
square lattice of size $L$. The walker  starts at a random initial site  
${\bf i}=(i_{x},i_{y})$, where $0 \leq i_{x},i_{y} \leq L-1$. 
At each  step the walker has four 
 possible moves, to one of the sites  
${\bf i_{R}}=(i_{x}+1,i_{y})$, 
 ${\bf i_{L}}=(i_{x}-1,i_{y})$, 
 ${\bf i_{U}}=(i_{x},i_{y}+1)$ and 
 ${\bf i_{D}}=(i_{x},i_{y}-1)$, 
each picked with equal probability. 
The boundaries are open, and thus a random walker starting at any site 
$(i_{x},i_{y})$ will eventually fall off the edge \cite{Walsh1981,Walsh1982}. 
 The number of moves it will  
make depends on the location of the initial site  as well as on the 
 particular realization of the  random moves generating the 
 path of the given walker. 
Therefore, there is a probability distribution 
 $p_{\bf i}(n)$, where $n=1,2\dots,\infty$, 
 that  a random walker starting at site $(i_{x},i_{y})$  
 will fall off the edge after $n$ moves. 
We will first calculate the average of this distribution given by  
\begin{equation} 
{ 
 \langle n_{\bf i} \rangle= \sum_{n=1}^{\infty}n \cdot p_{\bf i}(n). 
} 
\end{equation} 
The boundary conditions are given by  $\langle n_{\bf i} \rangle = 0$ 
for sites beyond the edge of the $L\times L $ lattice, namely those for which  
$i_{x} = -1$ or $L$, or $i_{y} = -1$ or $L$. 
Since on its way to the boundaries the walker must pass through at  
least one of the nearest neighbors of the site ${\bf i}$, there is a 
relation between $\langle n_{\bf i} \rangle $ and the corresponding 
averages for its nearest neighbors of the form  
 
\begin{equation} 
\langle n_{\bf i} \rangle=\frac{1}{4}
\left( \langle n_{\bf i_{R}} \rangle + \langle n_{\bf i_{L}} \rangle+ \\ 
\langle n_{\bf i_{D}} \rangle+ \langle n_{\bf i_{U}} \rangle \right)+1 . 
\end{equation} 
 
\noindent
This set  of $L^2$ coupled linear equations can be 
written in a matrix form as  
\begin{eqnarray}  
 \frac{1}{4}\Delta \langle {\bf n } \rangle    =  {\bf 1},  
\end{eqnarray} 

\noindent
where the matrix $\Delta$ is identical to  the  
toppling matrix  of the  
BTW model, given in Eq. ~(\ref{eq:delta}). 
The vector $\langle {\bf n} \rangle $, consists of the $L^2$ components 
$\langle { n } \rangle_{i} = \langle n_{\bf i} \rangle$, 
where ${\bf i}=(i_{x},i_{y})$ and  
$i=Li_{x}+i_{y}$ ($i=0,1,\dots,L^{2}-1$). 
The $L^{2}$-dimensional vector ${\bf 1}$ 
is given by ${\bf 1} = (1,1,\dots,1)$. 
 
In order to reduce the  
number of equations we shall use the symmetry properties of the square  
lattice, that has 
one  horizontal,  one vertical  and two diagonal  reflection axes.  
 Any two  sites with the same symmetry properties are 
 called sites of the same ``type''. 
Due to the symmetry it is sufficient 
to examine the sites in the triangle bounded 
by the vertical axis from the center 
upwards  and by the diagonal axis from the 
center to the upper - right corner 
(Fig.~\ref{fig:lattice}). 
In this triangle 
there is one site of each type. 
The number of sites in the triangle is 

\begin{equation}  
 N =  \frac{(L+1)(L+3)}{8},  
\end{equation} 

\noindent
when $L$ is  odd, and 

\begin{equation} 
N  = \frac{L(L+2)}{8}, 
\end{equation} 

\noindent
when $L$ is even. 
The N linear equations for $\langle n \rangle_{i}$  
are of the form :  

\begin{equation}\label{eq:general} 
\frac{1}{4} D \langle {\bf n} \rangle  =  {\bf 1},  
\end{equation} 

\noindent
where D is an $ N \times N$ matrix. The matrix elements of D are:
 
\begin{eqnarray} 
D_{i,j} = \left\{ \begin{array}{ll} 
 4-f(i,i)  &  \ \ i=j \\ 
-f(i,j)& \ \  i \neq j, 			 
\end{array} 
\right. 
\end{eqnarray} 

\noindent
where $f(i,j)$ is the number of sites of type $j$ that are nearest neighbors  
to a site of type $i$. 
For $L = 3$ 
Eq.~(\ref{eq:general}) takes the form
 
\begin{equation}
\left( 
\begin{array}{ccc} 
1 & -1 &  0  \  \\ 
 -\frac{1}{4} & 1 & -\frac{1}{2}  \\ 
0 & -\frac{1}{2} & 1 
\end{array} 
\right)\  \  
\left( 
\begin{array}{c} 
\langle n \rangle_{1}\\  \langle n \rangle_{2} \\ \langle n \rangle_{3}  
\end{array} 
\right) 
=  
\left( 
\begin{array}{c} 
 1 \\ 1 \\ 1 
\end{array}  
\right), 
\end{equation} 

\noindent
and its solution can be easily found to be  
$\langle {\bf n} \rangle = (4.5, \ 3.5, \ 2.75 )$. 
Now, having  explicit values for the $\langle n \rangle_{i}$'s, 
the average path length $\langle n \rangle$ of a random walker 
that starts at a random site on the $3 \times 3$ lattice is 

\begin{equation}
\langle n \rangle = \frac{\langle n_1 \rangle 
+ 4\langle n_2 \rangle+4\langle n_3 \rangle}{9} \ = \ 3.277..
\end{equation} 
 
The probability distribution $p_{\bf i}(n)$  of a walker  
starting  at site ${\bf i}$ to fall off the edge after $n$ moves  
can also be calculated \cite{Soler1982}. 
One can then average $p_{\bf i}(n)$  over all lattice sites and obtain 
the probability distribution $P_L(n)$,
$n=1,2,\dots$,
that a walker starting at a 
random site on the lattice will fall  
off the edge after $n$ steps. 
This probability is given by  

\begin{equation}\label{eq:pavrage} 
 P_L(n)=\frac{1}{L^2}\sum_{\bf i}p_{\bf i}(n). 
\end{equation} 

\noindent
Note that for the random walk model,
$P_L(n)$
is defined only for
$n \ge 1$,
because the random walker must make
at least one move in order to fall off
the edge.
The normalization condition will thus
take the form

\begin{equation}
\sum_{n=1}^{\infty} P_L(n) =1.
\end{equation}

The moments of this distribution are given by
 
\begin{equation}
\langle n^q \rangle_L = \sum_{n=1}^{\infty} n^q \cdot P_L(n) . 
\label{eq:np1} 
\end{equation} 

\noindent
The average path length (first moment) is given by Eq. 
(\ref{eq:np1})
with $q =1$, the second moment by
$q =2$ and so on.
The calculation of $p_{\bf i}(n)$ is done recursively starting from 
the boundaries. 
The probability that a walker starting at site ${\bf i}$ will fall  
off the edge after $n$ steps is given by  
 
\begin{eqnarray}\label{eq:dist} 
p_{\bf i}(n)=\frac{1}{4} \left[ p_{\bf i_{R}}(n-1)+p_{\bf i_{L}}(n-1)+ 
p_{\bf i_{U}}(n-1)+p_{\bf i_{D}}(n-1) \right], 
\end{eqnarray} 

\noindent
namely, it is the average over the four nearest neighbors, 
of the probabilities that a
walker starting in one of them will fall off the edge after
$n-1$ steps.
The boundary conditions are $p_{\bf i} (n) =0$ where 
$i_{x}=-1,L$ or $i_{y}=-1,L$ and $n=1,2,\dots, \infty$,
reflecting the fact that these indices represent sites
that are already over the edge.
The initial conditions for the recursive procedure for
the calculation of  
$p_{\bf i}(n)$, $n=1,2,\dots$,
are given by $p_{\bf i}(1) =1/4$  for all  
the edge sites, except for the corner sites, 
for which $p_{\bf i}(1) =1/2$. 
For all other sites $p_{\bf i}(1) =0$. 
The distribution $p_{\bf i}(n)$, $n=2,\dots$, 
is calculated recursively 
using Eq.~(\ref{eq:dist}) for all sites, 
starting at $n=2$ and increasing $n$ by $1$
after each scan of the lattice. 
The average path length $\langle n \rangle$ of a random walker starting at 
a random site is then obtained from Eqs.~(\ref{eq:pavrage}) 
and   ~(\ref{eq:np1}). 
 
\section{The connection between sandpile models and random walk models} 

Consider a grain deposited at a random site in the unrestricted  
two-state Manna  
model. It will initiate a small or a large avalanche and will typically  
stay on the lattice for many subsequent avalanches until it  
will fall off the edge. Most of this time the grain is alone in 
its site. Whenever it will share the site with another grain, 
both of them will topple randomly (and independently)  
to nearest neighbor sites. The path of the grain on the lattice is, 
in fact, a random walk since there is no correlation between 
one move to another
\cite{Dhar1999c}. 
This path starts at the site into which the grain was deposited
and ends at the edge site through which it leaves the system.
The different walkers are 
uncorrelated since the directions chosen by the two grains 
that topple from an unstable site
are independent.  
The only correlation is between the times that different walkers  
make their random moves. 
This temporal correlation appears because  two walkers need  
to occupy the same site in order to move. 

This property, that the path of each grain in the sandpile model
is a random walker, is not limited to the unrestricted two-state
Manna model. It is a general property of the conservative-Abelian
models, in which the grains are discrete entities. 
It is thus a common property of
models that belong to both the deterministic
and the stochastic universality classes. 
Consider, for example, the BTW model. 
It is convenient to consider
the grains as distinguishable particles by
naming each one of them
according to the running number 
representing the order
of their deposition into the system. 
When an unstable site topples, we will pick the grain that entered 
this site first, and choose randomly one of the four directions
($N, E, S, W$) for it to move. For the second grain we will choose
randomly among the three remaining directions, and so on for the
third and fourth grains. It is clear that there
is no correlation between the directions of consecutive moves 
of each grain and no bias. Therefore, each grain follows a path
of a random walker. Unlike the unrestricted two-state Manna model,
in the BTW model there is correlation between the directions of
different grains that topple from the same site, since they cannot
move in the same direction. 

Each avalanche in the sandpile models starts with a new grain 
deposited randomly. Therefore, on average each avalanche drops one grain off the  
edge. Since the grains follow random walker paths from the random
initial site to an edge site, 
the average number of hops that take place 
in a single avalanche must be equal to the average number 
of steps, that is required for a random walker deposited  
randomly on the lattice to reach the edge. 
We thus conclude that the average avalanche size of a sandpile
model on a lattice of size $L$ is the same for the BTW and Manna
models, and equal to the average path length of a random walker deposited 
randomly on the same lattice.

\section{Simulations and Results} 

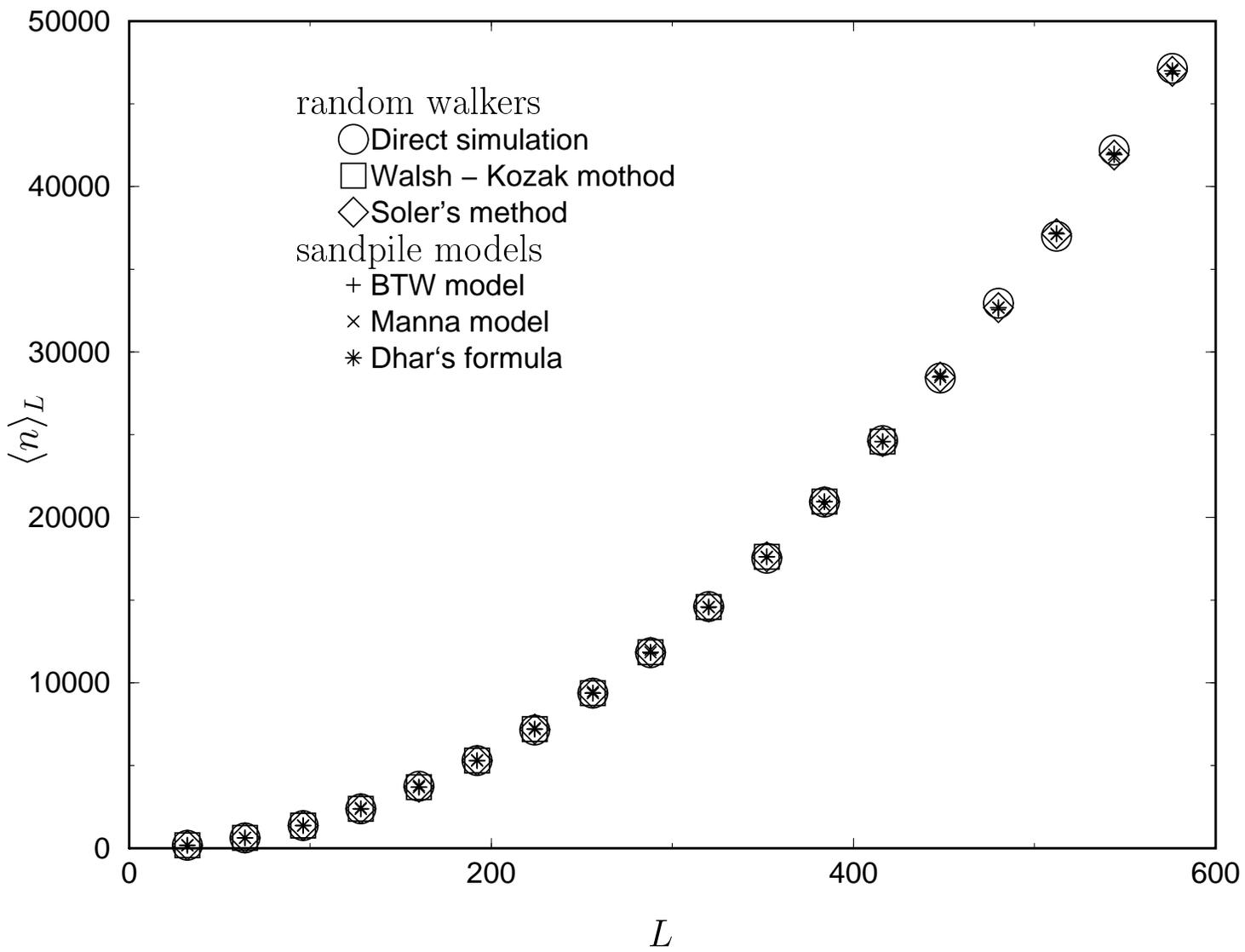
\begin{figure} 
\caption{ 
The average avalanche size for the BTW and Manna models,
and the average length of the random-walk paths (starting
from random sites) on a finite square lattice with open boundaries, 
vs. the lattice size $L$. 
For any value of $L$, it is found that the average 
avalanche sizes and the average
path lengths are all equal.
} 
\label{fig:averagen} 
\end{figure}

To examine the connection between the random walk model
and the sandpile models on finite lattices, we have performed numerical
simulations of both systems.
Direct simulations of the random walk model 
were performed on a square lattice of size $L$
with open boundaries. 
In each run the walker started at a random site on the lattice. 
The random walk path was generated until the walker fell off the
edge. 
The path length, namely the number, $n$, of moves it made from the
initial site to the edge was recorded. From this data, 
the distribution of path
lengths $P_L(n)$ was generated and its average
$\langle n \rangle_L$ 
was calculated.
The average   
$\langle n \rangle_L$ 
was also calculated using the
Walsh-Kozak method
\cite{Walsh1981,Walsh1982}.
The distribution
$P_L(n)$
was also calculated using the
Soler 
method
\cite{Soler1982}
and 
$\langle n \rangle_L$ 
was extracted from it.
Direct simulations of the BTW and Manna models were performed,
from which the avalanche size distributions were obtained. The
average avalanche size for each model was calculated as a function
of the lattice size $L$. 
The average avalanche size for the BTW model was also calculated
using Dhar's formula
[Eq.~(\ref{eq:Dhar1})]. 
The values of 
$\langle n \rangle$ vs. $L$,
obtained for both the random walk and sandpile models
are shown in 
Fig.~\ref{fig:averagen}.
They all coincide perfectly, except for some slight fluctuations
in the direct simulation data for the larger values of $L$.
This confirms the connection between the random walk and sandpile models.
Fitting 
$\langle n \rangle_L$ vs. $L$
to a polynomial function we obtain that
$\langle n \rangle_L = a L^2 + b L$,
where $a = 0.14$ and $b = 0.56$. 

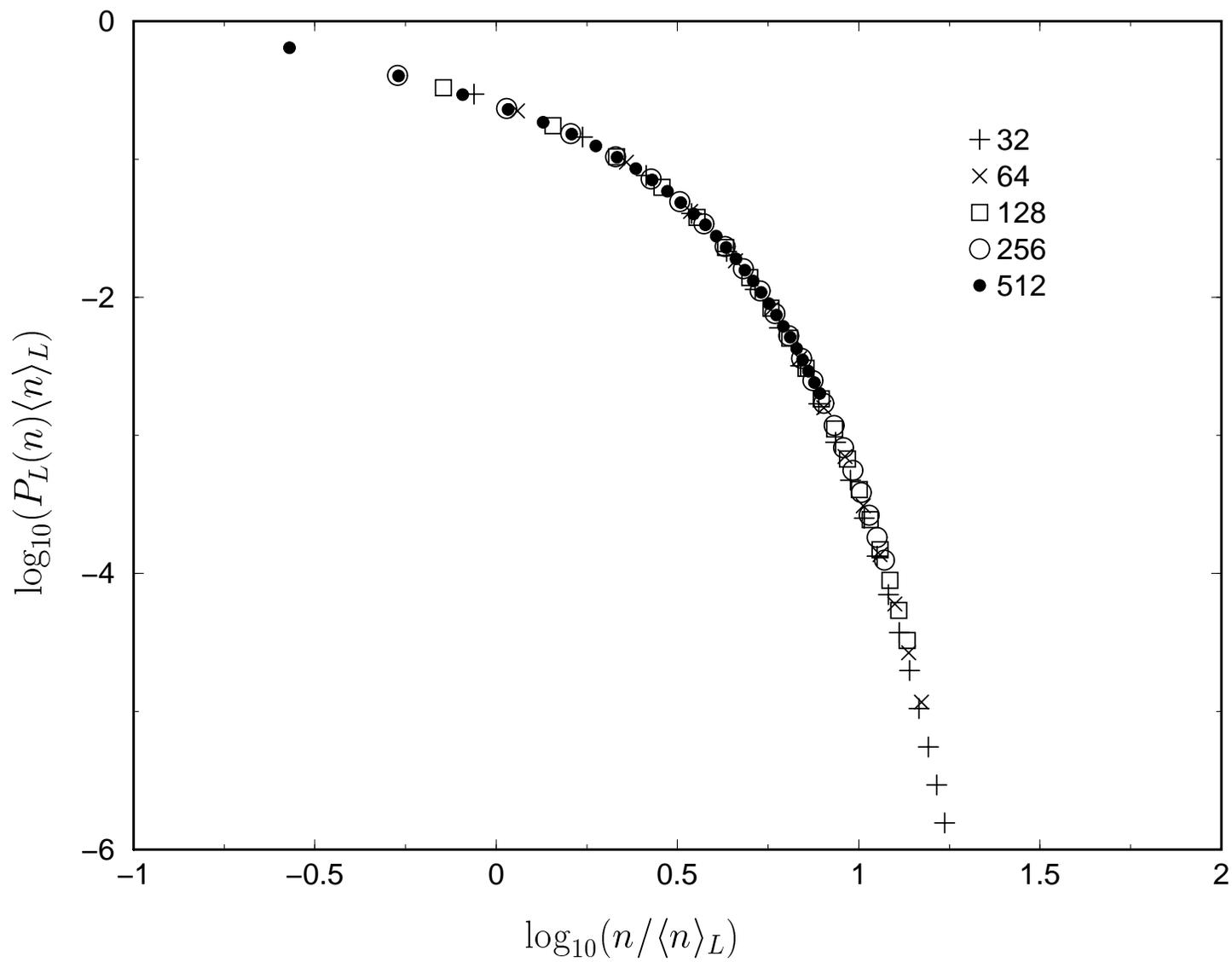
\begin{figure} 
\caption{ 
Scaling function of the path-length distribution of a random walk on
finite lattices of sizes 
$L = 32, 64, 128, 256$ and 512. 
Here, $n$ is the walker length,  
$\langle n \rangle_{L}$ 
is its average and $P(n)_{L}$ is the probability 
to obtain a path of length n on a lattice of size L. 
The five graphs coincide with each other. 
The slope in the linear range is 
$-1/2$, as expected for the random walk.
} 
\label{fig:scaling} 
\end{figure}

While the averages are found to be the same for the sandpile models
and the random walk models, the distributions 
$P_L(n)$
turn out to be 
different.     
To calculate the distribution 
$P_L(n)$
of path lengths of random walkers 
starting from random sites on a square lattice of size $L$,
we used the
Soler method
\cite{Soler1982}. 
The calculation was done for
square lattices of sizes  $L= 32, 64, 128, 256$ and 512.   
To obtain a scaling function we rescaled $n$ for each
system size by the average path length
$\langle n \rangle_{L}$.
The scaling function
$P(n/\langle n \rangle_L) = \langle n \rangle_{L} \cdot P_L(n)$
is shown in 
Fig.~\ref{fig:scaling}
on a double logarithmic scale.
The scaling function exhibits a linear range,
up to an upper cutoff around
$n / \langle n \rangle_{L} =1$.
The slope of the linear range 
turns out to be
approximately $-1/2$. 
This function can be considered  
in the framework of first passage
problems of a random walks on a
finite lattice
\cite{Montroll1964}.

The avalanche size distributions for the BTW and Manna
models were obtained from direct numerical simulations
for lattice sizes
$L = 32, 64, 128, 256$ and 512.  
The rescaled distribution functions are shown in
Fig.~\ref{fig:mannascale}(a) for the BTW model
and in
Fig.~\ref{fig:mannascale}(b) for the Manna model.
In both cases the data collapse is not complete, due
to the finite size dependence of the critical exponent
$\tau_L$.
Fitting the data for $L=512$ to 
Eq. ~(\ref{eq:power}) 
we obtain that
$ \tau_L = 1.12 \pm 0.02 $ 
for the BTW model
and
$ \tau_L = 1.27 \pm 0.02 $ 
for the Manna model,
in agreement with previous results
\cite{Manna1991a,Milshtein1998,Lubeck2000}.

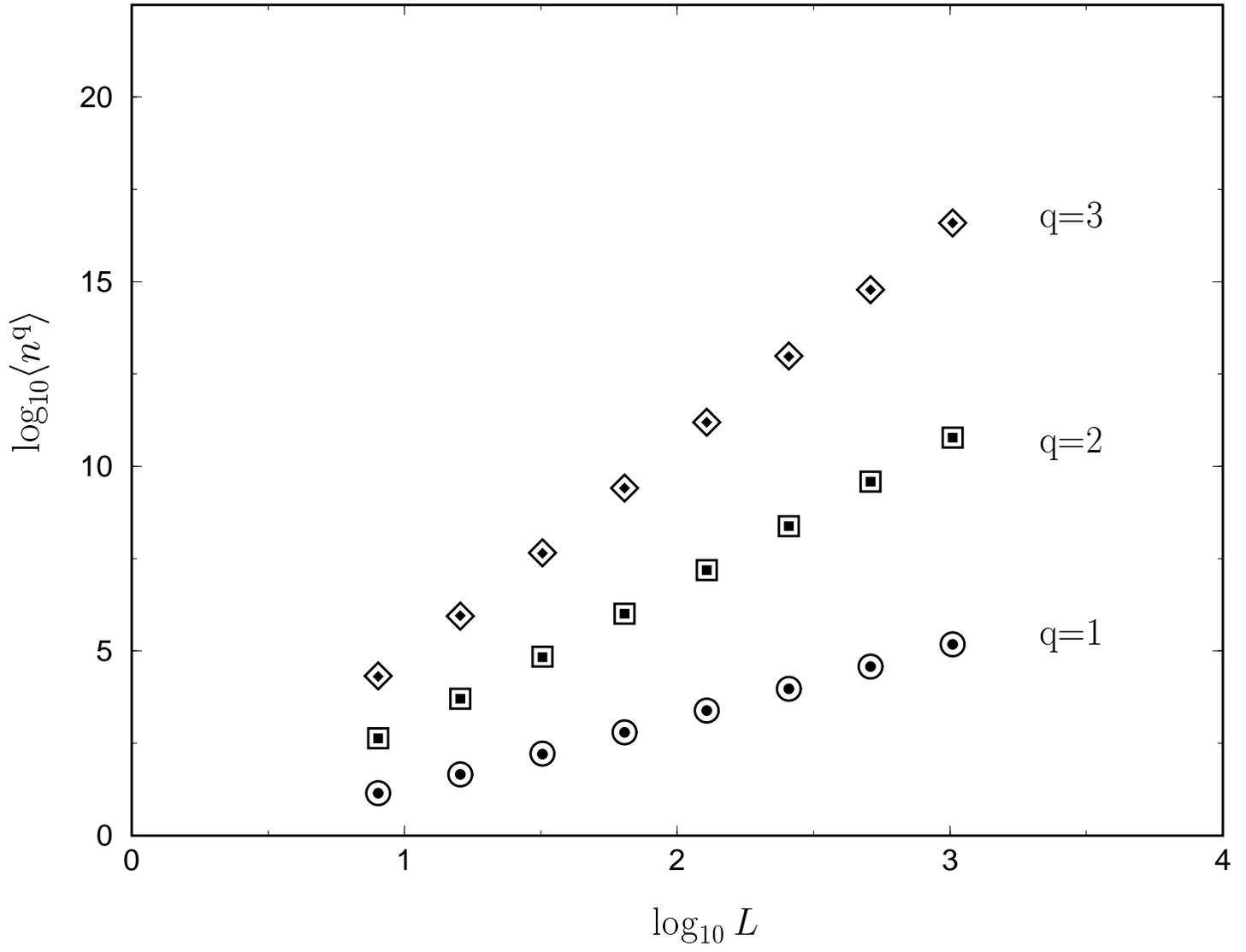
\begin{figure} 
\caption{ 
The first three moments of the distribution of the path-lengths of  
the random walk vs. system size. The filled symbols are the results of
direct simulations. The empty symbols are the results of calculations 
using the Soler method. 
The slopes of the best linear fits are 1.98 $\pm 0.04$, 3.97 $\pm 0.06$, 
and 5.96 $\pm 0.08$, for $q=1, 2$ 
and 3 respectively.
} 
\label{fig:walkermoments} 
\end{figure}

The  first three  
moments of the distribution of path-lengths of  
the random walk vs. $L$
are shown in 
In Fig.~\ref{fig:walkermoments}.
The results were obtained both by 
direct simulation 
and by calculating $P_L(n)$ 
using the Soler method
\cite{Soler1982}.
Fitting these graphs to power laws in $L$
we find that
$\langle n^q \rangle_L \sim L^{\rm 2q}$ 
(the slopes of the best linear fits are 
$1.98 \pm 0.04$, 
$ 3.97 \pm 0.06$, 
and 
$5.96\pm 0.08$ 
for $q=1$, $2$ and $3$, 
respectively). 

For sandpile models, in the large system limit,
the first moment of the avalanche size distribution
scales like
$\langle n \rangle_L \sim L^2$.
Higher moments are expected to scale like

\begin{equation}
\langle n^q \rangle_L \sim L^{\sigma(q)},
\label{eq:scalmomq}
\end{equation}

\noindent
with 
$\sigma(q) > 2 q$ 
for $q > 2$.
The first three moments of the 
avalanches size distributions of the BTW
and Manna models,
vs. $L$, are shown in  
Fig.~\ref{fig:mannamoments}. 
The slopes of the best linear  
fits for the BTW model  
are 
$\sigma(q) = 1.98 \pm 0.02$, 
$4.68 \pm 0.04$, 
and 
$7.52 \pm 0.08$  
for $q=1$, $2$ and $3$, 
respectively.   
The slopes of the best linear fits 
for the Manna model  
are 
$\sigma(q) = 1.97 \pm 0.02$, 
$4.73 \pm 0.04$, 
and 
$7.48 \pm 0.08$ 
for $q=1$, $ 2$ and $3$, 
respectively.
These linear fits were obtained for lattice sizes in the
range $64 \le L \le 1024$.
The results for the first moment
are identical (within the error bars) for the
two models, and coincide with the results 
for the random walk model, and thus confirm the conclusions of the
analysis above. 
Surprisingly, 
the values of $\sigma(q)$ 
for the BTW and Manna models 
are approximately the same (within the error bars), 
also for $q=2$ and 3.
This is in spite of the fact that the avalanche size distributions
of the two models are characterized by different exponents $\tau_L$.
This behavior has to do with deviations from the power-law behavior
near the upper cutoffs of the distributions.
The results for the 
higher moments  
are in agreement with those presented in 
Refs.
\cite{Lubeck2000,Chessa1999b}.  

\section{Discussion} 
 
Power-law distributions were observed in a wide variety of
natural systems as well as in economic systems, computer networks,
linguistics and other fields.
Some examples include the energy distribution between scales in
turbulence
\cite{Kolmogorov1962}, 
the distribution of earthquake magnitudes
\cite{Gutenberg1956}, 
the distribution of city populations
\cite{Zipf1949,Zanette1997},
the distribution of income and wealth
\cite{Pareto1897,Mandelbrot1951,Mandelbrot1961,Mandelbrot1963},
the distribution of the number of links pointing to sites in the
internet
\cite{Barabasi1999,Albert2002} 
and
the distribution of the frequency of appearance of words in texts
\cite{Zipf1949}. 
A common feature of such system is that they consist of a large number
of elementary degrees of freedom that interact with each other in
a complex way. Power-law distributions typically appear when these interactions
give rise to long-range correlations with no characteristic length-scale.

Consider a power-law distribution of the form

\begin{equation}
P_L(n) = A(L) \cdot  n^{-\tau_L},
\label{eq:fss}
\end{equation}

\noindent
limited to the range between 
$n_{\rm min}(L)$
and
$n_{\rm max}(L)$.
For simplicity we will assume that the
lower cutoff is fixed to
$n_{\rm min}(L)=1$.
As in the case of the sandpile models, we will assume that
the upper cutoff is limited by the system size, $L$,
and that
$n_{\rm max}(L) \rightarrow \infty$
when
$L \rightarrow \infty$.
The probability distribution $P_L(n)$ should satisfy
the normalization condition

\begin{equation}
\int_{n_{\rm min}(L)}^{n_{\rm max}(L)} P_L(n) {\rm d}n =1,
\end{equation}

\noindent
namely,
$A(L) = (1-\tau_L)/(n_{\rm max}^{1-\tau_L} - 1)$.
In order for $A(L)$ to 
converge to a finite nonzero value as 
$L \rightarrow \infty$,
the exponent $\tau_L$ must satisfy 
$\tau_L > 1$ in the infinite system limit.
The first moment of the distribution,

\begin{equation}
\langle n \rangle_L = \int_{n_{\rm min}(L)}^{n_{\rm max}(L)} n P_L(n) {\rm d}n, 
\end{equation}

\noindent
thus takes the form

\begin{equation}
\langle n \rangle_L = 
{ {(1-\tau_L)}  { [n_{\rm max}(L)^{2-\tau_L} - 1] } 
\over 
{(2-\tau_L) } { [n_{\rm max}(L)^{1-\tau_L} - 1] } }.
\end{equation}

\noindent
We observe that for $\tau_L > 2$ the first moment converges to
a finite value in the infinite system limit.
On the other hand, for $1 < \tau_L < 2$, 
the first moment, 
$\langle n \rangle_L$,
diverges
for $L \rightarrow \infty$.
We thus obtain a connection between the behavior of the
first moment of the distribution in the infinite system limit
and the range of values that the exponent $\tau_L$ can take.

For the sandpile models studied here the exponent $\tau_L$ is in
the range 
$1 < \tau_L <2$, 
and indeed, the average avalanche
size diverges according to
$\langle n \rangle = a L^2$
as $L \rightarrow \infty$.
The upper cutoff $n_{\rm max}$
can be expressed as a function of $L$ and $\tau_L$:

\begin{equation}
n_{\rm max} = 
\left[ {{a (2 -\tau_L)} \over {\tau_L -1}}  \right]^{1 \over {2-\tau_L}}
L^{2 \over {2-\tau_L}}.
\end{equation}

\noindent
Using this upper cutoff in the calculation of higher moments
we obtain that in the infinite system limit they will scale
according to Eq.
(\ref{eq:scalmomq})
with

\begin{equation}
\sigma (q) = { {2(q+1-\tau_L)} \over {2-\tau_L} }.
\label{eq:sigmaq}
\end{equation}

\noindent
Note that Eq.
(\ref{eq:sigmaq})
predicts significantly different values
of $\sigma(q)$
for the BTW and the Manna models, due to the difference in the
values of $\tau_L$ for the two models.
On the other hand, 
Fig.~\ref{fig:mannamoments}
shows nearly identical values of
$\sigma(q)$, $q=2,3$ for the two models. 
The fact that these two moments coincide seems to be due
to the deviations from power-law behavior near the upper-cutoffs.
The effect of these deviations is significant for high moments.

Recently, multifractal scaling was observed in the avalanche size
distribution of the BTW model
\cite{Menech1998,Tebaldi1999,Menech2000}.
This indicates that a finite size scaling analysis
of the form of 
Eq.~(\ref{eq:fss})
is not sufficient for describing the scaling behavior
of the BTW model, although it was found to apply in the
case of the Manna model
\cite{Menech1998,Tebaldi1999,Menech2000}.

Exponents $\tau_L$ in the range 
$1 < \tau_L < 2$, 
were observed empirically in the distribution of earthquake magnitudes.
Many other systems exhibit values of $\tau_L$ in the range
$2 < \tau_L < 3$. In these systems the first moment is kept
finite in the infinite system limit, while the 
second moment, that characterizes the fluctuations in the
system diverges. 
Consider, for example, a directed graph model describing an internet-like
network.
Each node in the graph has a fixed number, $r$,
of links pointing outwards to other nodes.
The graph is constructed such that the probability
of each node to receive links from newly
added nodes is proportional to the number of incoming
links that it already has.
For a network that reached a size of $L$ nodes,
this process generates a power-law distribution
of the number of incoming links among the nodes.
In the resulting network the total number of outgoing
links must be equal to the total number of incoming links.
Since each node has $r$ outgoing links,
the average number of incoming links per node must be
$\langle n \rangle_L =r$,
independent of the size $L$ of the network.
Since the first moment of the distribution is kept finite,
while the second moment diverges as $L \rightarrow \infty$,
the exponent $\tau_L$
must be in the range
$2 < \tau_L < 3$
in the infinite system limit.
The notable feature of the network system is that the 
average, 
$\langle n \rangle_L$,
of the power-law distribution of the incoming links is forced 
to remain constant and independent of the system size.
Systems that have this feature are common. 
Other examples include 
the distribution of the number of citations to scientific papers.
Each citation is a directed link from a newer paper to an older
one. While the distribution of  
the number of
outgoing links per paper
is narrow,
the distribution of incoming links is broad, and resembles a
power-law distribution.
Another example is the distributions of income and wealths
in western societies, that were found to exhibit power-law
behavior, at least in the high income sectors,
with exponents in the range $2 < \tau_L < 3$.
Here the argument is not as easy to make. However, one may
argue that the average of these distributions must be connected
to the average productivity per worker. This productivity 
remains finite when the size of the economy increases.

\section{Summary and Conclusions} 

Abelian sandpile models (both deterministic and stochastic)
and random walk models have been studied on
finite square lattices with open boundaries.
The avalanche size distributions of the sandpile models,
as well as the distributions of the lengths of the random-walk
paths were calculated using various methods.
It was shown that, due to the conservation laws, 
the averages $\langle n \rangle$
of the avalanche size distributions of the deterministic
and stochastic models are the same, and that they are
both equal to the average length of the random-walk paths
starting from random sites on the same lattice.

\acknowledgments
We thank O. Malcai for many useful discussions.
This work was supported by the US-Israel Binational Science Foundation,
under Grant No. 9800097. 
 

\newpage 
\clearpage

\begin{figure} 
\caption{ 
The square lattice of size $3 \times 3$, in which the sites are
grouped and indexed according to the symmetry of their positions. 
The indexing starts from the central site and outwards.
} 
\label{fig:lattice} 
\end{figure}

\begin{figure} 
\caption{ 
The rescaled avalanche size distributions for the 
BTW model (a) 
and for the 
Manna model (b)
for lattice 
sizes $L = 32, 64, 128, 256$ and 512. 
It is observed that 
for the BTW model
the slope  
exhibits some dependence on $L$.
The slope for best fit obtained by linear regression
for lattice size  $512$ corresponds to 
$\tau_L = 1.12 \pm 0.02$
for the BTW model,
and 
$\tau_L = 1.27 \pm 0.02 $ 
for the Manna model.
} 
\label{fig:mannascale} 
\end{figure}

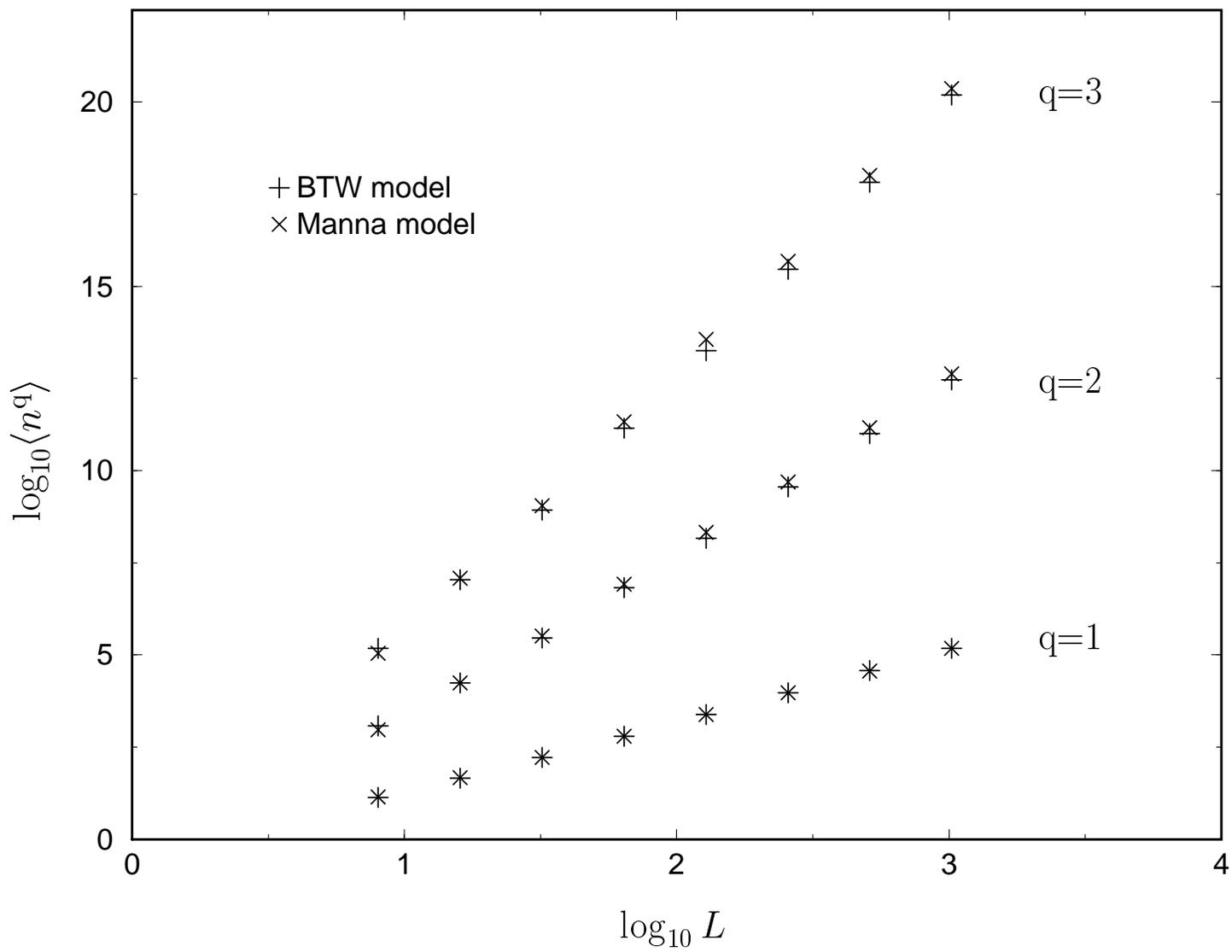
\begin{figure} 
\caption{ 
The first three moments of the avalanches sizes in the BTW and the Manna
models, vs. system size.  
The slopes of the best linear fits for the BTW model  
are $1.98 \pm 0.02$, $4.68 \pm 0.04$, and $7.52 \pm 0.08$ for $q=1, 2$ 
and 3 respectively.  
The slopes of the best linear fits for the Manna model  
are $1.97 \pm 0.02$, $4.73 \pm 0.04$, and $7.48 \pm 0.08$ for $q=1, 2$ 
and 3 respectively. 
The results for the first moment are the same for the two
models and coincide with the random walk model. The results for
higher moments of the two models are not identical, although the
differences are small. They both
are very different from the random walk results.
} 
\label{fig:mannamoments} 
\end{figure} 
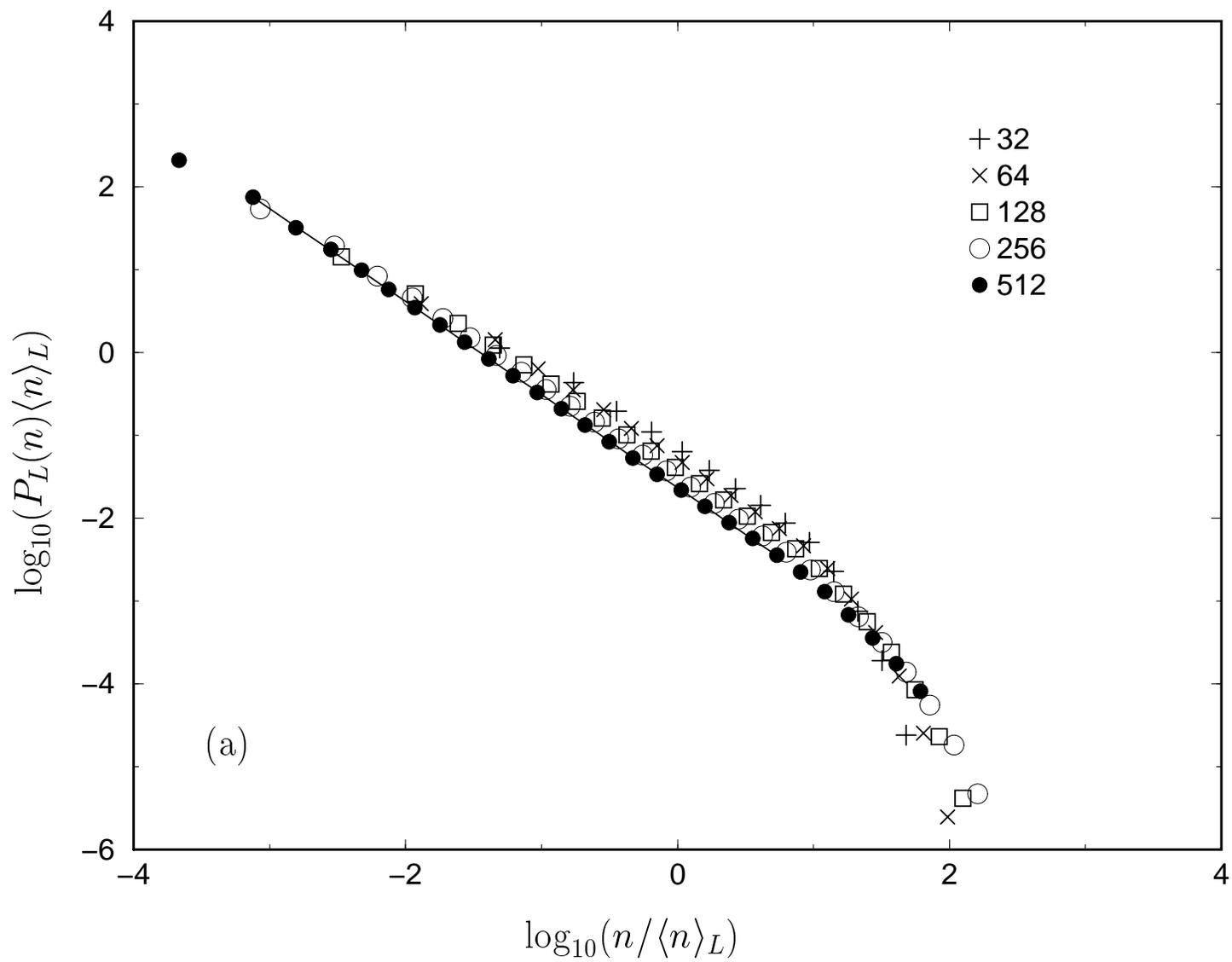
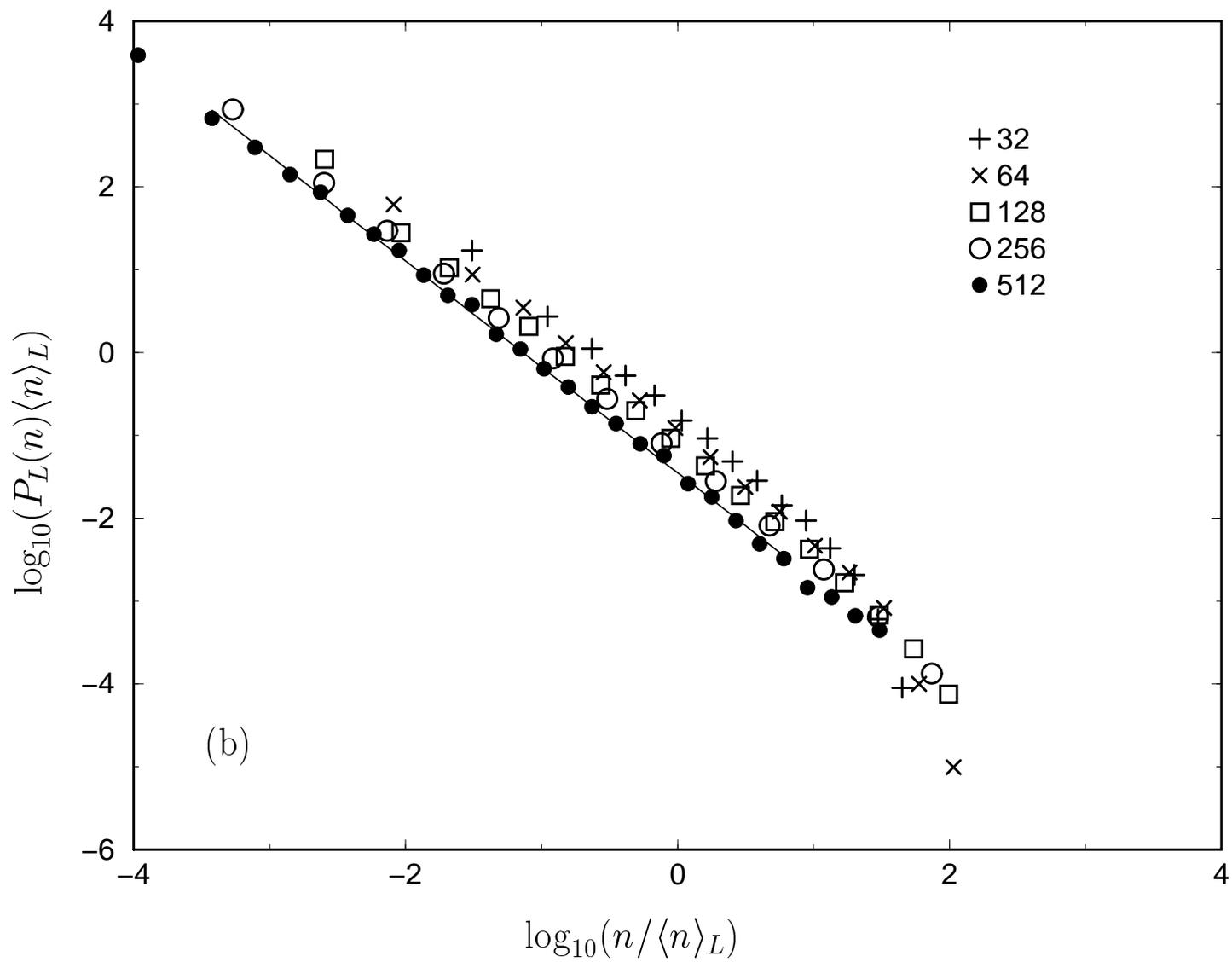
\end{document}